\documentstyle[prl,aps,preprint,tighten,floats]{revtex}
\begin{document}
\draft
\preprint{IASSNS-HEP-95/90, UPR-0684T, NSF-ITP-95-147, hep-ph/95111378
}
\date{November 1995}
\title{Implications of Abelian Extended Gauge
Structures From String Models}
\author{Mirjam Cveti\v c$^{3}$
\thanks{On sabbatic leave from the University of Pennsylvania.}
and Paul Langacker$^{1,2}$
}
\address{$^1$\ Department of Physics and Astronomy \\
          University of Pennsylvania, Philadelphia PA 19104-6396,\\
           $^2$ \ Institute for Theoretical Physics,\\
          University of California, Santa Barbara, CA 93106-4030,\\
           $^3$\  Institute for Advanced Study\\
           Olden lane, Princeton, NJ 08540\\ }
\maketitle
\begin{abstract}
Within a class of superstring vacua
which have an additional non-anomalous $U(1)'$ gauge factor,
we address the scale of  the $U(1)'$  symmetry breaking and
constraints on the exotic particle content and their masses.
We  also show that an extra gauge $U(1)'$ provides a new mechanism for
generating a naturally small effective $\mu$ term.
In general, existing models  are not consistent with  all
phenomenological constraints; however, they do provide
a testing ground to address the above  issues,
yielding a set of concrete scenarios.
Under the assumptions that the spontaneous   $U(1)'$ breaking takes
place  in the observable sector and that the  supersymmetry
breaking scalar mass square terms are positive at the
string scale, the breaking of    $U(1)'$  symmetry is radiative. It
can take place when the appropriate Yukawa
couplings of exotic particles are of order one, which occurs
for $Z_2\times Z_2$ fermionic  orbifold constructions at
symmetric  points of moduli space. The $Z'$  mass
is either of  ${\cal O}(M_Z)$,
when the symmetry breaking is due to a single standard model singlet,
or is  of a scale intermediate between the string and electro-weak scales,
determined by the
radiative corrections (or by competing non-renormalizable operators),
when the breaking is due to  two or more mirror-like singlets.
In the former case,
the  $M_{Z'}/M_Z$ hierarchy
achievable without excessive fine tuning is within future
experimental reach.
\end{abstract}
\pacs{}

\section{Introduction}
One of the challenges of string theory is to make a successful
connection  to the observable world. Attempts to  construct realistic
models have not been completely
successful  at
satisfying all  of the   phenomenological constraints.
The two major problems of  string models are:
\begin{itemize}
\item
Degeneracy of vacua: there are by now a large number of  string models,
  which in general  have a   large gauge symmetry
which is difficult to break to the standard model,
 a large number of families, and other exotic particles.
\item
Supersymmetry breaking: presently, we do not have a fully satisfactory
 scenario for supersymmetry breaking, either at the level
 of world-sheet dynamics or at the level of the effective theory.
\end{itemize}
Both problems are believed to have an ultimate resolution
 in the non-perturbative string dynamics.
However, in spite of the  absence of  a unique  string  vacuum
the string theory does provide certain generic and,
 for a  certain class of string vacua, specific predictions.

One of the  string predictions is the
 gauge coupling unification at
$M_{string} \sim g_U \times 5 \times 10^{17} $ GeV \cite{K},
 where $g_U$ is the gauge coupling at the string scale.
The observed couplings are approximately consistent with this
 prediction.
 Taking the observed $\alpha$ and weak
 angle $\sin^2 \theta_W$ as inputs
 and extrapolating assuming the particle content
of the minimal supersymmetric standard model (MSSM), one finds \cite{LP}
that the running $SU(2)$ and $U(1)$  couplings meet at a
scale $M_U \sim 3 \times 10^{16}$ GeV. One can then predict
$\alpha_s(M_Z) \sim 0.130 \pm 0.010$ for the strong coupling.
$M_U$ is about an order of magnitude below $M_{string}$. The
actual value of $\alpha_s(M_Z)$ is still controversial, with determinations
generally in the range 0.11--0.125 \cite{ALS}.
In any case, when properly viewed as predictions for $\ln(M_U/M_Z)$
and $1/\alpha_s$, the gauge unification works to within 10--15\%.

The  gauge group at $M_{string}$ is
generically not a grand unified gauge group, but a product of
semi-simple group factors. There are frequently additional factors
beyond the standard model group, such as extra $U(1)'$s.

There are a very large number of possible string models,
which correspond to consistent perturbative vacua of string theory.
It is not at present feasible to examine them all, but
one can take a less ambitious attitude and
consider only those  vacua which have the potential
to be realistic.
There are no known fully realistic
models, but nevertheless there are specific issues
that can be addressed  within  semi-realistic  models,
 which may in turn provide a pattern of general string model predictions.

In particular, we  study  a class of specific
 string vacua, which at $M_{string}$ possess $N=1$ supersymmetry,
  the standard  model (SM) gauge group as a part of the
 gauge structure, and a particle content that includes
 three SM families \cite{FIQS,NAHE,F} and  at least two SM
Higgs doublets, {\it i.e.},  the string vacua which
have at least the ingredients of the MSSM.
Specific models often contain additional matter multiplets
as well, often with exotic standard model quantum numbers.
In general, we do not want to give up
the approximate unification of gauge couplings.
This severely restricts
the possibilities for new exotic matter \cite{BL}. However, we will
sometimes relax the gauge unification constraint in discussing
specific models which illustrate interesting scenarios.

A number of such models
(not necessarily consistent with gauge unification)
were constructed
as orbifold models\cite{DHVW} with  Wilson
 lines, as well as models based on  the free (world-sheet)
 fermionic constructions\cite{ABK,NAHE,CHL}.
 The latter set of models are based on the $Z_2\times Z_2$ orbifold
 at special points of  toroidal and Wilson line moduli space.
 As for the issue of supersymmetry breaking
 one may parameterize  our ignorance
 by  introducing  soft supersymmetry breaking terms
 in the observable sector\footnote{For different scenarios,
 in which supersymmetry is broken in the moduli-dilaton sector,
 see Ref.\cite{BCI} and references therein.}.

A set of models of that type constitutes a starting point to address
 specific phenomenological issues. Here, we would like to derive
 the consequences  of  an enhanced gauge symmetry
 in the observable sector of the  above class of string vacua.
 For the sake of concreteness we shall concentrate
on an  additional non-anomalous $U(1)'$ symmetry. A generalization to more than
 one $U(1)$ symmetry factor is straightforward.

The phenomenology of heavy gauge bosons in gauge theories
has been extensively  studied in the past.
There are stringent limits on the mass and mixings of such
bosons from precision electro-weak experiments \cite{L} and
from direct searches  \cite{direct}.
The limits vary significantly from model to model
because of the different chiral couplings to the
ordinary fermions, but typically the mass of a
heavy $Z'$ must exceed $\sim$ 400 GeV, while the $Z-Z'$ mixing
angle must be smaller than a few times $10^{-3}$.

Furthermore, the
identification and diagnostic study of heavy gauge bosons
at future colliders has been investigated in
detail \cite{CG}.
There have also been studies of the present and future
constraints on possible exotic matter \cite{exotic}.
For example, some models predict the existence of a heavy
vector ($SU(2)_L$-singlet) charge $-1/3$ quark, $D_L-D_R$, which could
be produced at a hadron collider by ordinary QCD processes and
decay by $D_L-b_L,s_L,d_L$ mixing into, {\it e.g.,} $cW$, $bZ$,
or $bH$, where $H$ is a neutral Higgs boson. Currently,
$m_D > $ 85 GeV if it mixes mainly with $b$ \cite{exotic}.
However, heavy gauge bosons and exotic matter
have rarely been addressed together \cite{nrl}.
Additionally, on purely phenomenological grounds there is no particular
reason for the mass scale of new bosons or matter
to be in the window ({\it e.g.}, up to a few TeV) accessible
to present or future experiments.

In contrast, a  class of string models with the features mentioned above
 and an  additional $U(1)'$ symmetry
provide a  testing ground  to address  the  following
aspects  of an enhanced Abelian gauge symmetry:
\begin{itemize}
\item (i) a scenario, specifying the scale of $U(1)'$ symmetry breaking,
\item (ii) the mass scale and phenomenological implications of the exotic
particles associated with an enhanced gauge symmetry,
\item (iii) the implications of $U(1)'$ symmetry
for generating  a naturally small
effective $\mu $ term.
\end{itemize}

We have identified several distinct scenarios, each of which is
illustrated by
a specific model.  A thorough
analysis within a large class of such models awaits further investigation.

A major conclusion of this
paper is that  a large class of string models
considered here not only predict the existence of additional
gauge bosons and exotic matter particles, but  often imply the masses of
the new gauge bosons and  the exotic particles which necessarily accompany
them to be
in the electro-weak range.
Each specific model
leads  to  calculable  predictions (which, however, depend on
the assumed
soft supersymmetry breaking terms) for the masses, couplings, and
mixing with the $Z$ of the new boson(s), as well as for the
masses and quantum numbers of the associated exotic matter.

The paper is organized as follows. In Section 2 we specify
in more detail the features of the string models under investigation.
 In Section 3 we address specific $U(1)'$ symmetry breaking scenarios,
achievable without excessive fine-tuning of the soft supersymmetry breaking
parameters.
 We show examples in which the additional $Z'$ mass is: (a) comparable to
that of the $Z$ (already excluded); (b) in the  300 GeV to 1 TeV range,
which may be
still barely allowed but easily within the range of future or present
colliders; (c) at an intermediate scale ({\it e.g.}, $10^{8}-10^{14}$ GeV).
It is argued that in case (b) it is difficult though not impossible to
satisfy existing constraints on $Z-Z'$ mixing, especially for lower
values of $M_{Z'}$, and that $Z'$ masses above 1 TeV are not
expected (given our assumptions) without excessive fine tuning.
In Section 4
several issues  related to $Z'$ physics are briefly discussed.
In particular, it is argued
that in the case of $U(1)'$ symmetry breaking at the 100 GeV to 1 TeV scale it
may be possible to generate a naturally small effective $\mu$ term
by the vacuum expectation value of a standard model singlet field which
is charged under $U(1)'$.
On the other hand, in the absence of an
extended gauge symmetry (and assuming positive soft supersymmetry
breaking scalar
mass-square terms) standard model singlets will generally either not acquire
vacuum
expectation values, or will acquire VEV's at an intermediate scale. Such
singlets would not be suitable for generating an effective $\mu$ term, and
would yield intermediate scale masses for the exotic matter to which they
couple.
Conclusions are given in Section 5. In the Appendix the renormalization group
  equations  and the analytic solutions for the soft supersymmetry breaking
  mass-square terms are given  for specific  examples of Yukawa interactions
  under  specific assumptions for the soft supersymmetry breaking terms at
  $M_{string}$.

\section{General Features of a Class of String Models}

Let us first specify
the generic features of $N=1$ supersymmetric  string models with
the  standard model (SM) gauge group  $SU(2)_L\times U(1)_Y\times SU(3)_C$,
three ordinary families,  and at least two SM
doublets, {\it i.e.}, a set of models with at least
a particle content of the minimal supersymmetric standard model (MSSM).
Those models are primarily based on fermionic
$Z_2\times Z_2$ orbifold  constructions at a particular point
in the toroidal and Wilson line moduli space.

Such models  in general also contain a  non-Abelian ``shadow''
sector group and a  number of  additional $U(1)$'s, one of them
generically anomalous.  The  shadow sector is a set of
SM singlets that transform non-trivially under the
non-Abelian shadow gauge group. However, they  communicate to
the observable sector through additional $U(1)$ factors.
The shadow sector non-Abelian gauge group  may play a role in
dynamical supersymmetry breaking.
In addition, there are  a large number of additional matter multiplets,
which  transform non-trivially under the $U(1)$'s  and/or the  standard
model  symmetry.

Most of the available models of that type  correspond to the level
one Ka\v c-Moody algebra for the non-Abelian gauge
 group factors, which in turn ensures that  the
 chiral super-multiplets are all in the fundamental or singlet
  representations of the SM  and the  non-Abelian shadow sector gauge group.
In addition, from a set of models we  select only
 those with  $SU(3)_C$, $SU(2)_L$ and $U(1)_Y$
 all embedded into the $SU(5)$ gauge group, since for
  other types of embedding  the normalization of  the $U(1)_Y$ gauge group
  coupling is different from the one leading to the
   gauge coupling unification in the MSSM model.

The fact that at $M_{string}$ the observable sector gauge
group is  not $SU(5)$, but the SM gauge group,
implies \cite{S} that the theory in general contains
fractionally charged color singlets. A generic prediction for
fractionally charged color singlets may have important
phenomenological\cite{AADF} consequences.

Due to an  anomalous  $U(1)$  symmetry at  genus-one, there
is an additional contribution  of  ${\cal O}(M_{string}^2)$\cite{DSW,ADS}
 to the corresponding $D$
 term\footnote{In certain models there may be an additional
  numerical suppression factor of ${\cal O} (10^{-2})$, rendering
  the scale of the genus-one contribution to be
  smaller than $M_{string}$
  by  one to two orders of magnitude.}. The contribution of
  such a term is cancelled \cite{DKI,DSW,ADS} by
 giving nonzero  vacuum
  expectation values (VEV's)   of ${\cal O}(M_{string})$ to certain
  multiplets in such a way  that the  $D$ flatness and  $F$ flatness
  condition is maintained at genus-one level of the  effective string theory,
  thus providing a  mechanism for  `restabilizing' the vacuum at  genus-one.
  At the same time, the nonzero VEV's can  be chosen
  (at least in principle) in such a way  that, while the
  SM gauge group  remains intact,   a number of additional non-anomalous
$U(1)$'s
  are broken at $M_{string}$  as well. In addition, a number of  multiplets
become massive. Thus,  the enhanced  gauge symmetry and the
exotic particle content of  the observable sector
is in general drastically reduced. Nevertheless,
there are often one or more non-anomalous $U(1)$'s and associated exotic
matter that are left unbroken.
The study of symmetry breaking scenarios
of these leftover non-anomalous $U(1)'$
symmetries is the subject of this paper.

However, such models in general suffer from  one or more of the
following deficiencies: (i) It is not clear that for a desired gauge
group choice there always exists a choice of VEV's which
would ensure the $F$  flatness to all orders in the non-renormalizable
terms\footnote{In certain instances such constraints can be obtained
by applying selection rules for the corresponding string amplitudes, as
developed for orbifold \cite{CFTO} and blown-up orbifold compactifications
\cite{C}.  See also \cite{FIQS}.}. (ii) It is also not
 clear how the supersymmetry breaking
  scenario can be implemented. Namely, the gaugino condensation in the
 ``shadow'' sector  may not be possible, due to to a
 large  number of additional  shadow sector matter multiplets
 which may render the  shadow sector non-Abelian gauge group
 non-asymptotically free.
We shall not address the dynamical origin of  supersymmetry breaking --
in particular, the difficulties with  gaugino condensation
 or related issues of  dynamical symmetry breaking in  the
 shadow sector of the theory\footnote{See, however\cite{Nanopoulos,FH}.}.

There are also other,  phenomenological,  problems:
(i) In general the models have  additional color triplets
in the spectrum which could   mediate a too fast
proton-decay\cite {INQ,FPD}. (ii) The detailed  mass spectrum
of the ordinary   fermions\cite{FIQS,F}  is not realistic.
(iii) In the case of an enhanced symmetry in the observable sector
a scenario  for the enhanced symmetry breaking  may not be consistent with
phenomenological  constraints on the exotic multiplets,
such as gauge coupling unification.
(iv) There is no $\mu$ term in the  superpotential,
{\it i.e.},  the coupling between the two  SM  Higgs doublets is absent, since
these SM
doublets are massless at $M_{string}$. An effective $\mu$  term of  the
order of  soft supersymmetry breaking mass parameters is
needed for the low energy phenomenology \cite{ftp}.

Here, we shall concentrate on  phenomenological
consequences  of an additional non-anomalous $U(1)'$ symmetry.
We  shall not
attempt to solve all the problems
 of  such a class of  models, but rather address the  specific
  aspects of $U(1)'$ symmetry breaking.

\section{$U(1)'$ Symmetry Breaking Scenarios}

We shall confine the analysis to the set of models
whose features were specified in the previous section.
 In addition, we do not address dynamics associated with the
  shadow sector, and we parameterize the supersymmetry breaking
  by introducing
  soft supersymmetry breaking terms.
Within that context we are ignoring aspects of dynamical symmetry breaking due
the formation of condensates in the non-Abelian shadow sector,
 which could at the same time break the  additional $U(1)'$
at a large scale.

Under these assumptions
the $U(1)'$ symmetry breaking must take place
via the Higgs mechanism, in which the  scalar component(s)  of
 chiral super-multiplets $S_i$, which carry  non-zero charges
 under the $U(1)'$, acquire   non-zero vacuum expectation values (VEV)
and spontaneously break  $U(1)'$.
 The low energy effective action, responsible for
 spontaneous symmetry breaking (SSB), is  specified by the
  superpotential, K\"ahler potential,  and soft supersymmetry
breaking terms.

Assuming that the soft supersymmetry breaking mass-square terms for the
scalar fields are  positive
at $M_{string}$,
the only way of achieving SSB is via a radiative  mechanism.
Namely, since such SM  singlets  are  massless at $M_{string}$,
they  have no bilinear term in the superpotential, and
their soft supersymmetry breaking  mass-square terms  need to
be driven negative at lower energies to ensure  a
global minimum with nonzero VEV's for such fields. Since the soft
supersymmetry  breaking  mass-square terms are assumed  to be
positive (and often taken to be universal) at $M_{string}$,
the radiative breaking scenario can be achieved if
there are large Yukawa couplings of  the $S_i$'s
to a sufficient number of other fields.
This is most easily achieved if there are couplings to
 SM doublets or color triplets.
 In this case the  renormalization group equations
   ensure that the  corresponding mass-square terms for $S_i$
   can be driven negative\footnote{For the explicit form of  the
   renormalization group equations, which can be applied to this case,
    see, for example, Appendix A of Ref. \cite{CP}. See also \cite{mssmrad}.}.

Thus, the  scale of $U(1)'$ symmetry breaking depends  on
 both the  type of  SM singlets responsible for the $U(1)'$
 symmetry breaking  and on the Yukawa  couplings of
 such multiplet(s) to other exotic particles.
Interestingly, for the fermionic constructions based on $ Z_2\times Z_2$
 orbifolds at special points of toroidal and Wilson line moduli space,
the corresponding Yukawa couplings are  either of
${\cal O}(g)$ \footnote{There is still uncertainty about the precise
value of the Yukawa coupling, since the latter is related to
the  corresponding three-linear term in the superpotential  by a factor,
due to the K\" ahler potential contribution.
} or zero\footnote{The same is true
for the   Yukawa coupling of the ordinary families to the Higgs doublets.}.
Thus, if the relevant coupling is non-zero it
may be sufficiently large to  ensure radiative breaking
of the $U(1)'$ symmetry.

For  each of these possibilities the pattern of  $U(1)'$ symmetry breaking
and  the running of the gauge couplings still depends on  the  specific
exotic particle content and their couplings.
We shall now discuss the possible scenarios for observable sector
$U(1)'$ symmetry breaking.
For the sake of simplicity  we shall address  scenarios in which
the electro-weak symmetry is broken  due to the non-zero VEV of
the Higgs doublet  that couples to
the top-quark, {\it i.e.}, a large $\tan \beta$ scenario of the MSSM.
A generalization to scenarios that accommodate other ranges of $\tan \beta$
is straightforward. We will emphasize the general features which hold in each
scenario.
However, we emphasize that in each specific model the $Z'$ mass, mixing,
and couplings, as well as the properties of the exotic matter, are
in principle calculable, though in practice they depend on the details
of the soft supersymmetry breaking.

\subsection{$U(1)'$ Breaking without $U(1)'$ Charged   Standard Model Singlets}
\label{ones}
This is the case in which the low energy spectrum of the theory
contains  no  SM model singlets that are  charged under $U(1)'$,   or
in which their
effective mass-square terms  remain positive. In this case radiative breaking
of the electro-weak symmetry, due to
non-zero VEV of the  SM  Higgs doublet(s),
 would ensure  the SSB of the additional $U(1)'s$ as well,
 provided that they
are charged under $U(1)'$.

The $Z-Z'$ mass-square matrix is then of the form:
\begin{equation}
M^2_{Z-Z'} =
\left ( \matrix{\frac{1}{2} G^2H^2 &Gg' Q_H'H^2  \cr
Gg'Q_H'H^2& 2 g'^2{Q'}_H^2H^2} \right ),
\label{masscasea}
\end{equation}
where  $G\equiv\sqrt{g^2+g_Y^2}$. Here $g,g_Y,g'$ are the gauge couplings  at
the SSB scale for  $SU(2)_L$,
 $U(1)_Y$ and $U(1)'$, respectively,  $H$ is the VEV of the
  SM Higgs doublet(s), and $Q_H'$  is its $U(1)'$ charge.
  (There is an obvious generalization for the case of two Higgs doublets with
  different $U(1)'$ charges.)

In this case, the $Z'$ mass  is necessarily  of ${\cal O}(M_Z)$ and
 the $Z-Z'$ mixing is of $O(1)$.  Such a scenario  is likely
not to be  compatible with the current bounds on $M_{Z'}$\cite{L}.
An exception would be the (highly unlikely) possibility
that the  SM  Higgs doublet(s)
have $U(1)'$ charges  $|Q_H'| \gg 1$.  In this case
the mass of $M_{Z'}$ may be large enough and the $Z-Z'$ mixing
 sufficiently suppressed to evade the current bounds.

An illustration of this scenario is provided by  a
particular version of the
$Z_2\times Z_2$ orbifold model specified in Ref. \cite{F},  in which the
 vacuum at genus-one is restabilized by
 by giving nonzero VEV's to the SM singlets
$\Phi_{45,13},\Phi_{1,2}^{\pm},
 \bar \Phi_{1}^{\pm},\bar\Phi_{2,3}^-,\bar \Phi_{23,13},\xi_{1,2}$\footnote{
 For notation and quantum number assignments, see Ref. \cite{F}.}. In this case
there are no massless SM singlets that are charged under the surviving
$U(1)'$.
 The  SM Higgs doublet(s), responsible for the electro-weak SSB,
have $Q_H'=-1$.
The model contains the relevant Yukawa coupling between the third family
quarks and leptons to the SM Higgs, thus allowing for radiative
symmetry breaking.
Since  the $|Q_H'|$ charge is not large enough,
this SSB scenario is incompatible with
 current experimental bounds.
In addition the model has a number of additional light SM doublets  and color
triplets as well as fractionally  electrically charged  color singlets, which
affect the running of the gauge  couplings.

\subsection{Symmetry Breaking Due to One  $U(1)'$
Charged Standard Model Singlet}

Now suppose that the radiative
breaking of $U(1)'$ is due to one  SM singlet $S$.
Namely, only  $S$ has its effective mass-square  driven to a
negative
value in the infrared regime, thus allowing for a non-zero VEV.
The  $Z-Z'$ mass-square matrix is then:
\begin{equation}
M^2_{Z-Z'} =
\left ( \matrix{\frac{1}{2} G^2H^2 &Gg' Q'_HH^2  \cr
Gg'Q'_H H^2& 2 g'^2({Q'}_H^2H^2+{Q'}_S^2S^2)} \right ),
\label{masscaseb}
\end{equation}
where $H$ and $S$ now denote the  VEV's  for the SM Higgs doublet and singlet,
respectively, and $Q'_{H,S}$ are  the corresponding
$U(1)'$ charges.

The exotic matter to which $S$ couples acquires a mass of
order ${\cal  H}S$, where $\cal H$ is the relevant Yukawa coupling  between
the particular exotic  matter and  $S$.  In general, there
will be an additional  soft supersymmetry breaking mass term contributing to
the mass of the exotic matter even in the absence of the relevant Yukawa
coupling(s).

The nature of the $Z-Z'$ hierarchy now crucially depends on
 of the allowed VEV's $S$ and $H$, which are constrained by the
form of the  potential, and can be written for the  particular direction with
non-zero VEV's as:
\begin{equation}
V=-m_H^2H^2-m_S^2S^2+{1\over 8}G^2H^4+{g'^2\over 2} (Q_H'H^2+Q_S'S^2)^2,
\label{potential}
\end{equation}
where we have assumed that $m_{H,S}^2>0$.

One encounters the following two scenarios:
\begin{itemize}
\item (i): The relative signs of $Q'_H$ and $Q'_S$ are opposite.

 In this case the minimum of the
potential is for:
\begin{equation}
H^2={{4(m_H^2+{{|Q'_H|}\over {|Q'_S|}}m_S^2)}\over {G^2}}\ \ ,\ \
S^2={{m_S^2}\over
{|Q_S'|^2g'^2}}+{{|Q'_H|H^2}\over {|Q'_S|}}
\label{VEVS}
\end{equation}
and the $Z-Z'$ mass-square matrix is
\begin{equation}
M^2_{Z-Z'} = 2
\left ( \matrix{(m_H^2+{{|Q'_H|}\over {|Q'_S|}}m_S^2)
&{{2g'Q'_H }\over G} (m_H^2+{{|Q'_H|}\over {|Q'_S|}}m_S^2) \cr
{{2g'Q'_H}\over G}
(m_H^2+{{|Q'_H|}\over {|Q'_S|}}m_S^2)& {{4g'^2{Q'}_H^2}\over{G^2}}
(1+{{|Q'_S|}\over{|Q'_H|}})(m_H^2+{{|Q'_H|}\over
{|Q'_S|}}m_S^2)+m_S^2
} \right ).
\label{masscaseBa}
\end{equation}
It is difficult to achieve the needed
hierarchy between  $M_Z$ and $M_{Z'}$,
unless $|Q'_S| \gg |Q'_H|$ and $m_S^2 \gg m_H^2$, in such a way that
$|Q_H'/Q_S'|= {\cal O} (m_H^2/m_S^2)$.
The first condition is not normally expected to hold, except in the
limiting case $Q_H' = 0$.

\item (ii):  The relative signs of $Q'_H$ and $Q'_S$ are  the same.

The minimum of the potential (\ref{potential}) now occurs for:
\begin{equation}
H^2={{4(m_H^2-{{|Q'_H|}\over {|Q'_S|}}m_S^2)}\over {G^2}}\ \ ,\ \
S^2={{m_S^2}\over
{|Q_S'|^2g'^2}}-{{|Q'_H|H^2}\over {|Q'_S|}}
\label{VEVSSAME}
\end{equation}
and the $Z-Z'$ mass-square matrix becomes:
\begin{equation}
M^2_{Z-Z'} =  2\left ( \matrix{(m_H^2-{{|Q'_H|}\over {|Q'_S|}}m_S^2)
&{{2g' Q'_H} \over G}(m_H^2-{{|Q'_H|}\over {|Q'_S|}}m_S^2) \cr
{{2g' Q'_H} \over G}(m_H^2-{{|Q'_H|}\over {|Q'_S|}}m_S^2)&
{{4g'^2{Q'}_H^2}\over{G^2}}
(1-{{|Q'_S|}\over{|Q'_H|}})(m_H^2-{{|Q'_H|}\over {|Q'_S|}}m_S^2)+m_S^2
} \right ).\label{masscaseBb}
\end{equation}
In this case one encounters an interesting possibility for achieving a
hierarchy without an unusually small ratio of $ |Q_H'/Q_S'|$,
 provided $0<m_H^2-{{|Q'_H|}\over {|Q'_S|}}m_S^2
 \ll m_S^2$.
 In this limit, the $Z-Z'$ mixing angle is
\begin{equation}
\theta_{Z-Z'} \sim {{2 g' Q'_H} \over{G}}{{M_Z^2} \over {M_{Z'}^2}}.
\end{equation}
For small $g' Q'_H/G$
the $Z-Z'$ mixing could be sufficiently suppressed to  be consistent
with the experimental bounds for  $M_Z'\le {\cal O}(1)$ TeV.

\end{itemize}

One can also illustrate the above scenarios in a particular
 class of $Z_2\times Z_2$ models  discussed in Ref. \cite{F}.

Case (i): In this case the vacuum at genus-one is restabilized
by nonzero VEV's of the following SM singlets :
$\Phi_{23},\bar\Phi_{23}$,  while the following SM
model singlets should
necessarily have zero VEV's: $\xi_1,\xi_2,\xi_3$.\footnote{We have checked
 the
F-flatness  of these and subsequent choices of VEV's  only at
the level of renormalizable terms.}
  The  relevant  SM singlet $S$ is identified with the
 field $H_{18}$  with  $Q_S'=5/4$, and thus has  the {\it opposite} sign from
 $Q_H'=-1$.  In addition,  $H_{18}$ has a   Yukawa  coupling of ${\cal
O}(1)$ to two
 color triplets (in the notation of Ref. \cite{F}, the coupling is
 of the type $D_{45}H_{18}H_{21}$), and thus
 its mass-square can become negative in the infrared regime (see Appendix).
  The additional
 $U(1)'$ charged  SM singlets, {\it e.g.}, $H_{17}$, have mass-square terms
which
 remain positive  in the infrared regime.
However, in this case the  magnitude of charges $Q'_{H,S}$ is such that
 {\it no hierarchy} is possible, and thus the model is excluded by
 experiment.

 Case (ii): The vacuum at genus-one is restabilized
with non-zero VEV's
 for the SM singlets $\xi_1,\xi_3, \Phi_{23}$,
while $\xi_2,\Phi_{12},\bar\Phi_{23}$ should have zero VEV's.
$S$ is identified with the
 field $H_{17}$, which has  $Q_S'=-5/4$ and has a
 Yukawa  coupling of ${\cal O}(1)$ to two SM doublets
  (in the notation of Ref. \cite{F}, the coupling is
 of the type $h_2H_{16}H_{17}$). Thus,
its mass-square  can become negative in the infrared regime (see Appendix).
The additional
 $U(1)'$ charged  SM singlets, {\it e.g.}, $H_{18}$, again
 have mass-square terms which
 remain positive  in the infrared regime.

This  is the case in which a reasonable hierarchy can be achieved  without
fine tuning of the  soft supersymmetry breaking parameters.
In the case of simplified assumptions for  the  soft supersymmetry breaking
parameters (see Appendix), {\it i.e.}, only  universal
soft supersymmetry breaking mass-square terms are assumed to be non-zero,
the model yields
the mass parameters
 $m_H^2=0.40m_{3/2}^2,m_S^2=0.25m_{3/2}^2,
 m_H^2-m_S^2|Q'_H|/|Q'_S|=0.20m_{3/2}^2$, which unfortunately do not yield the
 necessary hierarchy.  However, a  minor deviation from  the above
assumptions for  the soft supersymmetry breaking parameters, or just an
example of a model with  a slightly different ratio of $|Q'_H|/Q'_S|$,   can
provide for a  hierarchy
$m_H^2-m_S^2|Q'_H|/|Q'_S|\ll m_{S}^2$, say,
$m_H^2-m_S^2|Q'_H|/|Q'_S|\sim m_S^2/10$. If, in addition, one
takes\footnote{This ratio depends on the  exotic particle content, which
may,   in addition to the  particles  contributing to the radiative breaking,
include additional exotic particles. The full exotic particle
content depends on the non-zero VEV's of the SM singlets restabilizing the
vacuum at genus-one.
In the above analysis we specified the
minimal choice of non-zero and zero VEV's of the corresponding  SM singlets,
in order to ensure radiative breaking of  $U(1)'$. In general, one also has to
ensure that the exotic particle content is compatible with the  unification of
the SM gauge coupling constants. This imposes another stringent constraint on
the  allowed exotic particle content.},
{\it e.g.,} $g' |Q'_H|/G\sim 1/4$,
one  obtains
$M_Z'^2\sim 10 M_Z^2$ and the mixing angle $\theta_{Z-Z'}\sim  0.05$.
In this example, $M_{Z'}$
is barely within the current experimental bounds,
while $\theta_{Z-Z'}$ is too large for most choices of $Z'$
couplings \cite{L,direct}. Somewhat larger values of $M_{Z'}$ and
smaller values of  $g' |Q'_H|/G$ may be consistent with observations.

This particular scenario is most interesting, since it in
principle allows, without excessive fine tuning of the
soft supersymmetry breaking parameters, for prediction of $M_{Z'}$ within
 experimental reach of present or future colliders. However,
when the  experimental bounds on $M_{Z'}$  exceed the 1 TeV region, this
scenario {\it cannot} be implemented without excessive fine tuning of the
soft supersymmetry breaking parameters or unusual
choices of the  $U(1)'$ charge assignments.

\subsection{Symmetry Breaking Due to Mirror-like Pairs of $U(1)'$
Charged Standard Model Singlets}

\label{flatsection}
In this case, negative mass-square terms are induced for two (or more)
$U(1)'$ charged SM  singlets $S_{1,2}$, whose $Q_{S_1,S_2}'$ charges have
opposite sign. One has flatness of the $D$ term along the
direction $Q'_{S_1}S_1^2=-Q'_{S_2}S_2^2\equiv S^2$.  One now has
to include the renormalization group improved potential, which
 along the flat direction is of  the form :
\begin{equation}
V=  m_S^2(\mu=S)S^2
\label{flatpot}
\end{equation}
Thus, the minimum occurs near the scale $\mu_{\rm crit}$ at which
$m_S^2$ turns negative.
In the case of radiative breaking with Yukawa couplings of ${\cal O}(1)$,
it turns out that  $m_S^2({\mu_{\rm crit}})$ is
much larger than the soft supersymmetry breaking mass terms.
For the
examples in the Appendix,  $\mu_{\rm crit}$ is typically
four to ten
orders of magnitude below $M_{string}$. Therefore, in the case of flat
directions the  scale of symmetry breaking, {\it i.e.,} the VEV of $S$,
is  ${\cal O}(10^{-10}-10^{-4})
M_{string}={\cal O}(10^{8}-10^{14})$ GeV.

Non-renormalizable terms  in the superpotential of
 the form $S^{K+3}/M_{planck}^{K}$  ($K\ge 1$) could in principle compete
 with the radiative corrections included in (\ref{flatpot}) and determine
  the scale to  be of the order of ${\cal O}([M_ZM_{planck}^K]^{1\over{K+1}})$
  \footnote{Such a scenario, with the interplay of the
   non-renormalizable superpotential terms and the negative
   mass-square terms, has been implemented in a related context
   of breaking of $E_6$ symmetry for Calabi-Yau compactifications
\cite{DMKN}.}.
For example, for
$K=1$  the symmetry breaking scale is of the order $10^{11}$ GeV.

 In either case the exotic  SM non-singlets  acquire mass
of the order of the intermediate scale.
On the other hand, it is straightforward to show that the mass of the
physical Higgs boson associated with $S$ is of the order of
the soft supersymmetry breaking  mass terms. The exotic matter
 to which $S$ couples via the  Yukawa couplings  of magnitude ${\cal H}$
acquires  mass of order ${\cal  H}S$,  {\it i.e.}, that of the intermediate
scale.   In the absence of the relevant Yukawa couplings, the
exotic matter would have a mass set
by soft supersymmetry breaking mass terms, and  thus
of the weak scale.

An illustration of this scenario of symmetry breaking is again provided by  a
version of the $Z_2\times Z_2$ orbifold model \cite{F},
 in which the  vacuum at genus-one is restabilized
 by giving nonzero VEV's to  the  SM singlets
$\Phi_{23}$ and
$\bar{\Phi}_{45}$, while the   SM model singlets $\xi_1,\xi_2,\xi_3$
should
necessarily have zero VEV's.   The  relevant  mirror-pair
SM singlets
 $S_{1,2}$  are identified with the
 fields $H_{17,18}$, respectively, with charges $Q_S'=\mp 5/4$.
  In addition, $H_{17}$ has a
 Yukawa  coupling of ${\cal O}(1)$ to two SM doublets
  ( of the type $h_2H_{16}H_{17}$), while  $H_{18}$ has a
  coupling of ${\cal O}(1)$ to two
 color triplets (of the type $D_{45}H_{18}H_{21}$), and thus
 its mass-square term can become negative for both fields in the infrared
regime
(see Appendix). Again,  additional $U(1)'$ charged  SM singlets,
{\it e.g.}, $H_{25}$, have mass-square terms which remain positive in the
infrared
regime.

\section{Other Implications of $U(1)'$}

In this section we briefly discuss other consequences of an
extended $U(1)'$ gauge symmetry.

One of the generic problems of the MSSM is the  so-called $\mu$-problem
\cite{muprob}. In the usual scenario for radiative electro-weak
breaking, the renormalizable terms in the superpotential, relevant
to the Higgs mechanism, are
\begin{equation}
  W= \mu h_1 h_2 + {\cal H}_t Q_LQ_Rh_1,
  \label{super}
\end{equation}
where $h_1$ and $h_2$ are the two  SM Higgs doublets which give
mass to the $t$ and $b$ quarks, respectively.
The quark doublet $Q_L$ and singlet
$Q_R$ are identified with the third family, and ${\cal H}_t$ is the
(large-${\cal O}(1)$) Yukawa coupling.

$h_1$ and $h_2$ are assumed to
acquire positive mass-square terms of   ${\cal O}(m_{3/2}^2)$ at a large scale.
The mass-square  term  for $h_1$ is driven negative at low energies
due  to the large Yukawa coupling ${\cal H}_t$.
However, to achieve a realistic  mass spectrum,
the  bilinear $\mu$ term is needed,
which yields supersymmetric mass contributions for $h_i$ and contributes
to the chargino and neutralino masses.  Also, the
soft supersymmetry breaking term  of the type $B\mu h_1 h_2$, where $B$ is a
soft supersymmetry breaking parameter,
leads to $h_1-h_2$ mixing and to a nonzero VEV for $h_2$, yielding the
$b$ quark mass.

For a realistic mass spectrum  $\mu$ in (\ref{super}) should be
${\cal O}(m_{3/2}^2)$. However,  since it is a coupling in the
supersymmetric Lagrangian, then, at least in the context of the MSSM,
 there is no reason for it not be much larger. This problem is referred to as
the $\mu$-problem

In the next to minimal supersymmetric model (NMSSM)\cite{nmssm}, the
$\mu$-problem is addressed in the following way.
One assumes that the $\mu$ term is absent due to a symmetry,
and that it is replaced by an effective $\mu$ given by the VEV of
of a standard model singlet $S$.
That is, one replaces the superpotential (\ref{super}) by
\begin{equation}
  W = S h_1 h_2 + {\cal H}_t Q_LQ_Rh_1 - {\kappa\over 3} S^3.
  \label{supernmssm}
\end{equation}
It is assumed that the scalar component of $S$ acquires a  non-zero VEV
due to a negative soft supersymmetry breaking mass-square term of ${\cal
O}(m_{3/2}^2)$ at the
at the electro-weak scale.
The cubic $\kappa$ term in (\ref{supernmssm}) yields a quartic
term for $S$ in the scalar potential, so that for
$\kappa$ of  ${\cal O}(1)$, the VEV of $S$ is of ${\cal
O}(m_{3/2}^2)$, thus yielding an effective $\mu$ parameter of ${\cal
O}(m_{3/2}^2)$.

Another possibility is that $\mu$ is absent in the superpotential
due to a symmetry, but that an effective $\mu$ term is generated by
non-renormalizable operators \cite{gm,munro}.  In particular,
terms  in the K\" ahler potential, which are proportional to
$ h_1 h_2$,
are transmitted to the observable sector as effective $\mu$ terms, due to
gravitational effects  after the spontaneous supersymmetry breaking,
and are therefore of the same order of magnitude
as the soft supersymmetry breaking terms \cite{gm}.

Remarkably, in string models
 $\mu = 0$ at $M_{string}$, since by definition $h_1$ and $h_2$ belong
to the massless sector and do not have bilinears in the
superpotential.  It was shown  in Refs. \cite{KLM} that a
number of string models possess  non-renormalizable terms in the K\"ahler
potential, which are proportional to $ h_1 h_2$, and may thus provide
a resolution to the $\mu$-problem \` a la \cite{gm}.

On the other hand,  the  NMSSM  mechanism is difficult
to implement directly in string models. There may be an  additional
singlet (or singlets) $S$ which has the appropriate coupling
$S h_1 h_2$, and $S$ may acquire a negative mass square at
the electro-weak scale if it has additional Yukawa couplings
to exotic matter.
However, the underlying symmetries
generally forbid the appearance of the needed ${\kappa\over 3} S^3$ term
in the superpotential, so there is no quartic term in $S$
in the potential. The situation is analogous to that of
a $D$-flat direction discussed in Section \ref{flatsection}.
In particular, one expects
$S $ to acquire an intermediate scale VEV either due to higher order terms
in the effective potential (when $m_S^2$ goes through 0), or
by non-renormalizable terms in the superpotential. In either case,
the VEV of $S$, and thus the scale of electro-weak breaking,
would be many orders of magnitude above
$m_{3/2}^2$, which is clearly not satisfactory.

The situation changes in the presence of an additional non-anomalous $U(1)'$
gauge symmetry, \footnote{For scenarios addressing $\mu$-problem within
anomalous
$U(1)'$ see Ref. \cite{JS}.} provided the term $ h_1 h_2$ is not a gauge
singlet\footnote{This
is not the case for the models in \cite{F}, but is expected in, {\it e.g.,}
models for which the $U(1)'$ is associated with the embedding of
$SO(10)$ into $E_6$.}.  Then, an $S h_1 h_2$ term can generate
an effective $\mu$ that is naturally of order $m_{3/2}$, provided
$S$ develops a VEV by the radiative mechanism.
The soft supersymmetry breaking term $A S h_1 h_2$ generates the
needed $h_1 - h_2$ mixing.

For example, suppose the
superpotential contains
\begin{equation}
  W \sim S h_1 h_2 + {\cal H}_t Q_LQ_Rh_1 + {\cal H}_E E_1E_2S,
  \label{superexz}
\end{equation}
where $E_{1,2}$ represent additional exotic matter. Assuming
that the  corresponding soft supersymmetry
 breaking mass-square terms $m_S^2$, $m_{h_1}^2$, and $m_{h_2}^2$
are positive at $M_{string}$,
then  for sufficiently
large Yukawa couplings ${\cal H}_{t,E}$ and a sufficiently large
representation of  $E_i$ under the SM gauge group,
both $m_S^2$ and $m_{h_1}^2$ are
driven negative at low energies. Two examples, in which
the $E_i$ are respectively $SU(2)$ doublets and $SU(3)$ singlets,
are described in the Appendix.
As discussed above, without  $U(1)'$ symmetry  the VEV of $S$ will be at an
unacceptably large intermediate scale. However, if $S$ carries
a nonzero charge under an extended gauge symmetry, then the $U(1)'$
$D$ term will provide, in the
 the absence of flat
 directions (scenarios discussed in subsection \ref{ones}),
a quartic potential for $S$, ensuring
that the VEV  $S$ will be of order $m_{3/2}$.

Thus, the existence of an additional
gauge $U(1)'$ provides a scenario which
leads to an effective $\mu$ term of ${\cal O}(m_{3/2})$,
thus providing a resolution to the
$\mu$-problem within string models.  This mechanism is
complementary to the  mechanism \cite{KLM}  based on non-renormalizable terms
\cite{gm,munro}.
The latter  mechanisms can occur if
there is no  additional
$U(1)'$ symmetry. However, if such a $U(1)'$ is present it will
forbid the needed terms  proportional to  $h_1 h_2$
in the K\" ahler potential or higher order
terms in the superpotential.

Let us briefly comment on a few related topics.
Even in the absence of a $U(1)'$ gauge symmetry it is possible
for SM singlet scalar fields to acquire negative mass-square values at
low energies due to the radiative mechanism if they have sufficiently large
Yukawa couplings to other fields. As was discussed above in
connection with the $\mu$-problem, such scalars generally
will not have quartic terms in the potential, and thus they
would acquire intermediate scale VEV's.

We have seen that under a certain set of assumptions
the VEV's of standard model scalars will typically be either
at the electro-weak scale, if there is an addition $U(1)'$ with
no dangerous flat directions, or at an intermediate scale.
Intermediate scales are of interest in implementing the
seesaw model of neutrino mass. However, one may still need non-renormalizable
terms in the superpotential to implement realistic neutrino
 mass scenarios \cite{CL}. Also, one
promising scenario for baryogenesis \cite{fy} is that a large
lepton asymmetry is generated by the decays of the heavy Majorana
neutrino associated with the seesaw, and then converted to a baryon
asymmetry during the electro-weak transition.
Such scenarios, while very attractive, cannot occur if
there is a $U(1)'$ which is only broken at the electro-weak to TeV
scale \cite{buch}, unless, of course, the heavy Majorana neutrino is
a $U(1)'$ singlet.

The radiative mechanisms discussed in this paper require the
existence of sufficiently large Yukawa couplings to drive the mass square
values
of the SM singlet $S$ negative. This is most easily implemented if there
exists exotic matter which transforms non-trivially under the SM gauge
group. The exotic matter will then acquire mass terms given by the relevant
Yukawa coupling times the VEV of $S$, as well as contributions from
soft supersymmetry breaking.
Such exotic matter typically exists in string models.
However, if it carries SM quantum numbers it can destroy the
success of the gauge coupling unification \cite{BL}. Such effects largely
cancel if the light exotic matter corresponds to complete $SU(5)$
multiplets, but that is not typically expected in the types of
semi-realistic models we are discussing.
There are many ways in which cancellations between different
multiplets or with other effects, such as higher Ka\v c-Moody levels,
can occur. However, preserving gauge unification without fine-tuning
is a stringent constraint on string model building, with or without
an additional $U(1)'$.

\section{Conclusions}

We have explored the possible scenarios
for (non-anomalous) $U(1)'$ symmetry breaking, as is expected
for a class of  string compactifications with the
standard model gauge group and additional $U(1)$ gauge factors.
This is the case for a large number of string vacua based on
 orbifold models \cite{DHVW} or based on  the free
 fermionic constructions\cite{ABK,NAHE,CHL}.
 Under the assumption that the symmetry breaking takes place in the observable
 sector and that the soft supersymmetry breaking scalar mass-square terms
 are positive, the
 breaking is necessarily radiative and
 requires the existence
 of additional matter, most easily associated with
 standard model  non-singlets.   Then, within a particular model
 with definite soft supersymmetry breaking terms, the
 symmetry breaking pattern, the couplings and the masses of the
  new gauge bosons, and those of the
  accompanying exotic particles are  calculable.
In that sense the string models yield  predictions for the new physics
associated with the new gauge bosons.

It turns out that for the class of  string models considered  the radiative
 $U(1)'$ symmetry breaking
is either of    ${\cal O}(M_Z)$  or  the
intermediate scale of order  $\ge 10^{8-14}$ GeV. However, in both cases the
mass of the associated physical Higgs bosons is in the electro-weak region.
Our  major conclusion, therefore, is that  a large class of string models
considered here not only predict the existence of additional
gauge bosons and exotic matter, but   often imply that their masses should be
in  the electro-weak range.
Many such models are already excluded by indirect or direct constraints
on heavy $Z'$ bosons, and the $Z-Z'$ mixing is often too large, especially
for lower values of $M_{Z'}$.
The scenario in which  $M_{Z'}$ is in the
electro-weak range  allows, without excessive fine tuning of the
soft supersymmetry breaking parameters, for predictions of $M_{Z'}$ within
 experimental reach of present or future colliders.
On the other hand,  when the  experimental bounds on $M_{Z'}$  exceed the 1
TeV region, this
scenario {\it cannot} be implemented without excessive
fine tuning of soft supersymmetry breaking parameters, or unusual
choices of  $U(1)'$ charge assignments.

In addition,  $U(1)'$ symmetry, broken at the electro-weak scale, provides a
simple mechanism  for generating an effective  $\mu$-term of the order of the
electro-weak scale.

 The analysis  has set the  stage for detailed investigation
 of the potentially phenomenologically viable string models with additional
$U(1)'$ gauge symmetry.

\acknowledgments
The work was supported in part by U.S. Department of Energy Grant No.
DOE-EY-76-02-3071, the National Science Foundation Grant No. PHY94-07194, the
Institute for Advanced Study funds  and J. Seward
Johnson foundation (M.C.),
and the National Science  Foundation Career Advancement Award
PHY95-12732 (M.C.). We  would like to thank
P. Binetruy, M. K. Gaillard, A. Faraggi, G. Kane, V. Kaplunovsky, J. Louis,
 J. Polchinski, and  especially  S. Chaudhuri for discussions. M.C.
acknowledges
 hospitality of the Institute for Theoretical Physics, Santa Barbara, where the
 work was initiated.

\newpage
\appendix{\bf Appendix }

We provide analytic expressions for the renormalization group equations
(RGE's)  for the running of the soft supersymmetry breaking mass-square terms
in the presence of sufficiently large Yukawa couplings.

The analysis is based on rough approximations,  which, however, illustrate the
mechanism of radiative supersymmetry breaking for different types of Yukawa
couplings of the $U(1)'$ charged  SM singlet $S$ to the SM $SU(2)$
doublets $h_{1,2}$
and or $SU(3)$ triplets $D_{1,2}$.
In the RGE's for the
soft mass-square terms   the following approximations are used:
we  replace the   Yukawa couplings  with a constant value close to the
infrared fixed point, and the three-linear  soft supersymmetry breaking terms
as
well as the  gaugino masses are  assumed to  be smaller than the soft
supersymmetry breaking mass-square terms. The first approximation is in
general a good one, since the Yukawa couplings, which are ${\cal O}(1)$ at
$M_{string}$,  reach  the infrared fixed point, which is  also ${\cal O}(1)$,
quickly. On the other hand, neglecting the
three-linear soft supersymmetry breaking
terms and gaugino masses may not be a good approximation in
general. However, the qualitative picture is not drastically changed with
the inclusion of such terms\footnote{See  Appendix B  of Ref. \cite{CP}
 for a related discussion of analytic expressions within the
 MSSM  with a heavy
 fourth family, when such terms are included.}.

 For simplicity we shall also
 assume universal positive mass-square terms $m^2_{3/2}$
 at $M_{string}$ and take the infrared fixed point value of the
 corresponding Yukawa coupling to be 1. For
 different  infrared fixed point  values ${\cal H}_Y$ of the corresponding
 Yukawa couplings,  the parameter $t$ in the analytic expressions
 is replaced
 by ${\cal H}_Y^2t$.

 We discuss the results for the following
 types of Yukawa interactions:
 \begin{itemize}
 \item
 Yukawa coupling: $h_1h_2S$.

 In this case the RGE's assume the form:
 \begin{equation}
2{ {dm_{h_{1,2}}^2}\over{{dt}}} =
 {{dm_S^2}\over{dt}} = 4\Sigma,
\label{doubletRGE}
\end{equation}
where $t={1\over{16\pi^2}}\log (\mu/M_{string})$ and
$\Sigma=m_{h_1}^2+m_{h_2}^2+m_S^2$.

 At scale $\mu$ the solution is
 \begin{equation}
 m^2_{h_1,h_2}={1\over 4}m_{3/2}^2 +{3\over 4}m_{3/2}^2{\rm e}^{8t}\ ,\
 m_S^2=-{1\over 2}m_{3/2}^2 +{3\over 2}m_{3/2}^2{\rm e}^{8t}\ ,
 \label{sol1}
 \end{equation}

 For $\mu={\cal O} (100)$ GeV, ${\rm e}^{8t}\sim 0.17$
 and thus $m_S^2$ is negative. $m_S^2$ goes through
 zero at $\mu_{\rm crit} \sim 2 \times 10^8$ GeV.

\item
 Yukawa couplings $D_1D_2S$.

The RGE's are
 \begin{equation}
3{ {dm_{D_{1,2}}^2}\over{{dt}}} =
 {{dm_S^2}\over{dt}} = 6\Sigma,
\label{doubletRGE2}
\end{equation}
where
$\Sigma=m_{D_1}^2+m_{D_2}^2+m_S^2$.

 At scale $\mu$
 \begin{equation}
 m^2_{D_1,D_2}={2\over 5}m_{3/2}^2 +{3\over 5}m_{3/2}^2{\rm e}^{10t}\ ,\
 m_S^2=-{4\over 5}m_{3/2}^2 +{9\over 5}m_{3/2}^2{\rm e}^{10t}\ .
 \label{sol2}
 \end{equation}

 For $\mu={\cal O} (100)$ GeV, ${\rm e}^{10t}\sim 0.11$
 and thus $m_S^2$  approaches
 a negative value, a factor $\sim 2$ more negative than the one in the
 previous case.
 In this case,
$\mu_{\rm crit} \sim  10^{12}$ GeV.

 \item  Yukawa coupling $Q_LQ_Rh_1$.

 This coupling is relevant to the breaking of the electro-weak
 gauge symmetry by the Yukawa coupling of
 the SM Higgs doublet $h_1$ to
the  left-handed  quark doublet $Q_L$  and the
 right-handed quark  $Q_R$, which are of course to be identified
 with the third family.

The RGE's are
  \begin{equation}
3{ {dm_{Q_L}^2}\over{{dt}}} =
 {3\over 2} {{dm_{Q_R}^2}\over{dt}}   =
  {{dm_{h_1}^2}\over{dt}}=6\Sigma_1,
 \label{ewrad}
\end{equation}
where
$\Sigma_1=m_{Q_L}^2+m_{Q_R}^2+m_{h_1}^2$.

One obtains
 \begin{equation}
 m^2_{h_1}=-{1\over 2}m_{3/2}^2 +{3\over 2}m_{3/2}^2{\rm e}^{12t}.
 \label{solew}
 \end{equation}

 \item  Yukawa coupling $Q_LQ_Rh_1+h_1h_2S$.

 This coupling is relevant when the same
 SM Higgs doublet couples to both the  SM singlet $S$ as well as through
 another Yukawa coupling  to the  left-handed  quark doublet $Q_L$  and the
 right-handed quark  $Q_R$.

 In this case the RGE's are
  \begin{eqnarray}
2{ {dm_{Q_L}^2}\over{{dt}}}&=&
 {{dm_{Q_R}^2}\over{dt}}   =  4\Sigma_1\nonumber\\
2{ {dm_{h_2}^2}\over{{dt}}}&=&
  {{dm_S^2}\over{dt}} \ = \ 4\Sigma_2\nonumber\\
  {{dm_{h_1}^2}\over{dt}}&=&6\Sigma_1+2\Sigma_2
 \label{doubletRGE3}
\end{eqnarray}
where
$\Sigma_1=m_{Q_L}^2+m_{Q_R}^2+m_{h_1}^2$ and
 $\Sigma_2=m_{h_1}^2+m_{h_2}^2+m_S^2$.

The solution  for $m_{h_1}^2$ and $m_S^2$ is
 \begin{equation}
 m^2_{h_1}=-{5\over 7}m_{3/2}^2 +{12\over 7}m_{3/2}^2{\rm e}^{14t}\ ,\
 m_S^2={1\over 7}m_{3/2}^2 +{6\over 7}m_{3/2}^2{\rm e}^{14t}\ ,
 \label{sol3}
 \end{equation}
indicating that at $\mu={\cal O}(100)$ GeV  the Higgs doublet $h_1$
  acquires negative mass-square terms, while the SM singlet $S$ has a
  mass-square term that remains positive.

 \item  Yukawa coupling $Q_LQ_Rh_1+h_1h_2S+E_1E_2S$.

This example is relevant to the generation of an effective
$\mu$ term. $h_1$ has the normal Yukawa coupling to the
third generation of quarks, the second term will play the
role of a $\mu$ term when $S$ acquires a VEV, and the
$E_1E_2S$ coupling to exotic particles $E_1$ and $E_2$ can
drive $m_S^2$ negative at an intermediate scale. The resulting set
of coupled RGE's can easily be solved numerically. We find that
in the case that $E_i$ represent a second pair of $SU(2)$
doublets both $m_S^2$ and $m_{h_1}^2$ are indeed negative
(and are equal within the approximations), and that at
$\mu={\cal O}(100)$ GeV
 \begin{equation}
 m_S^2 = m^2_{h_1}= -0.52 m_{3/2}^2\ , \
     m^2_{h_2}= +0.77 m_{3/2}^2,
 \label{solm1}
 \end{equation}
 thus allowing the generation of an effective $\mu$ term.
Similarly, when the $E_i$ are $SU(3)$ triplets,
 \begin{equation}
 m_S^2 = -0.74 m_{3/2}^2\ , \
 m^2_{h_1}= -0.48 m_{3/2}^2\ , \
     m^2_{h_2}= +0.84 m_{3/2}^2
 \label{solm2}
 \end{equation}
at $\mu={\cal O}(100)$ GeV.

 \end{itemize}

\end{document}